\documentclass{appolb}
\usepackage{graphicx}
\usepackage{epsfig}

\begin{document}
\pagestyle{plain}
\title{INFLUENCE OF CHARMING PENGUINS ON THE EXTRACTION OF  $\gamma$ IN $B \rightarrow PP$ DECAYS.}
\author{M. Sowa
\address{The Henryk Niewodnicza\'nski Institute of Nuclear Physics\\ 
Polish Academy of Sciences\\
 31-342 Krak\'ow, Poland\\
}}
\maketitle

\begin{abstract}
Charmless $B \rightarrow PP$ decays are studied using  flavor SU(3) symmetry.
Amplitude with charming penguin topology is considered for two cases: with zero and with arbitrary strong phase.
Two sets of data (an older and the most recent one) are used in the fits, so that the stability of the fits is
 tested. It is shown that within the present uncertainties in the data the
 parameters of the fit may be significantly modified, especially the $\gamma$ angle.
 The fits indicate the strong phase of the charming penguin amplitude to be around $\pm 20^{o}$.
\end{abstract}

{\small PACS numbers: 13.25.Hw, 12.15.Hh}

\section{Introduction}

$B$-meson decays are an excellent source of information about the CKM mechanism and allow us to test our
 understanding of the CP violation. In nonleptonic $B$
decays we must deal with final states interactions (FSI) as well, since they may modify the values of the extracted
 parameters. It is hard to take FSI into consideration properly since
 there are a lot of possible decay channels.
\\

During the recent years several authors have investigated various possible corrections
 due to FSI. Most of the analyses take into account the elastic and inelastic effects
 arising from intermediate states containing light quarks ($u$, $d$, $s$) [1-6] and apply
  symmetries of strong interactions (isospin, SU(3)) to reduce the number of parameters.
  Some authors argued that intermediate states containing charmed quarks ($c$) may also play an important role [7-13].
\\

In the present paper we analyse $B$ decays  into two light noncharmed pseudoscalar mesons.
 We consider FSI originating only from the intermediate states containing c quarks.
In sections 2 and 3 we introduce our parametrisation and relations between the amplitudes.
 Section 4 contains the description of the fit procedure and our results together with the CP
 asymmetry predictions. Finally, a short summary is given in section 5.

\section{Short-distance amplitudes}

 The decays of $B$ meson into two noncharmed pseudoscalar mesons are characterised by 10 $SU(3)_{f}$ invariant
 amplitudes corresponding to the specific quark-line diagrams. As in \cite{ZL,ZL2} we use four
 dominant amplitudes: tree $T(T^{'})$, colour-suppressed $C(C^{'})$, penguin
 $P(P^{'})$ and singlet penguin $S^{'}$. Unprimed (primed) amplitudes denote
 strangeness conserving (violating) processes and are related to each other.
 Topological decompositions of decay amplitudes can be found in \cite{ZL}.
\\

 We use the Wolfenstein parameters: $\lambda=0.222$,
 $A=0.832$, $\bar{\rho}=0.224$ and $\bar{\eta}=0.317$ \cite{fl}.
 All relations bellow are calculated up to $O(\lambda^4)$ unless explicitly written
 otherwise. Terms proportional to $\lambda^4$ are kept on account of complex factor in $P^{'}$, which may interfere with FSI correction. We assume that all short-distance (SD) strong phases
 are negligible. For the tree amplitude we have
\begin{equation}
T^{'} = \frac{V_{us}}{V_{ud}} \frac{f_{K}}{f_{\pi}} T =0.278 T,
\end{equation}
where $\frac{f_{K}}{f_{\pi}}$ is the SU(3) breaking factor. Both $T$ and $T^{'}$
amplitudes have a weak phase equal $\gamma$.
We assume that SD penguin amplitudes are dominated by $t$ quark contribution.
When terms of order $\lambda^{4}$ are included, the strangeness violating penguin amplitude $P^{'}$ acquires a small weak
phase. Thus, $P^{'}$ can be represented as a sum of two terms, the second one due to the $O(\lambda^4)$ correction:
\begin{equation}
P^{'} = \frac{V_{ts}}{V_{td}} P = -(5.241+0.105 e^{i\gamma} )|P|
\end{equation}
Penguin amplitude $P$ has weak phase -$\beta$. We used in our fits the value of $\beta=24^{o}$ consistent with the world average.
The singlet penguin has the same phase as penguin $P'$:
\begin{equation}
S^{'}=e^{i arg(P')}|S'|
\end{equation}
\\

Finally, we accept relations between the tree and colour-suppressed amplitudes:
\begin{equation}
C=\xi T,
\end{equation}

\begin{equation}
C^{'}=(\xi -(1+\xi )\delta_{EW} e^{-i\gamma})T^{'}
\end{equation}
where $\xi=0.17$  and $\delta_{EW}=0.65$. The last equation includes electroweak penguin
$P_{EW}^{'}$. The EW penguin contribution $\sim \delta_{EW} e^{-i\gamma}$ was calculated (see e.g. \cite{Neu2,Neu:98})
without $\lambda^{4}$ corrections. This fact should not affect the fits much  since $P_{EW}^{'} \sim S^{'}$ \cite{h} and the small correction
 in $S^{'}$ is
practically invisible in the fits (the only changes we
observed were in the asymmetry for the $B^{+} \to \eta^{'} K^{+}$ decay channel).

\section{Long-distance charming penguins }
It was argued [7-13] that the intermediate states composed of charmed mesons ($D\bar{D}$, etc.), generated from the
$b \rightarrow c \bar{c} d(s) $ tree amplitudes $T_{c}^{(')}$, may lead via rescattering to amplitudes of  
penguin topology with an internal $c$ quark (the "charming penguin"). Our calculations are similar as in
the case of long-distance $u$-type penguins \cite{ZL2}. Assuming SU(3) symmetry, we can redefine
penguins:
\begin{equation}
P^{(')} \rightarrow P^{(')}+id_{c}T_{c}^{(')}
\end{equation}
where $d_{c}$ is related to the size of the LD charming penguin and is a complex number in general.
Because we do not have information about $d_{c}$ (or $T_{c}^{(')}$), it is convenient to
introduce the following parametrisation:
\begin{equation}
id_{c}T_{c}^{(')}=P^{(')}_{cLD}e^{i \delta_{c}}
\end{equation}
Strong phase $\delta_{c}$ and size $P^{(')}_{cLD}$ of the charming penguin are additional free
parameters
in our fits. The weak phases are determined by the tree amplitudes $T^{(')}_{c}$ and are either $\pi$ or 0.
We can eliminate $P^{'}_{cLD}$ using the relation
\begin{equation}
\frac{P^{'}_{cLD}}{P_{cLD}}=\frac{T^{'}_{c}}{T_{c}}=\frac{V_{cs}}{V_{cd}}=-4.388
\end{equation}
\\

Short-distance charming penguin $P_{c}^{'}$ has the same weak phase as $P_{cLD}^{'}$.
It can be included in a new redefined charming penguin
\begin{equation}
P^{(')}_{cef}e^{i \delta} = P^{(')}_{c}+P^{(')}_{cLD}e^{i \delta_{c}}
\end{equation}
with new effective size and strong phase.

\section{Results of fits }
We minimise function f defined as:
\begin{equation}
f =\sum_i{\frac{(B_i^{\rm theor}-B_i^{\rm exp})^2}{(\Delta B_i^{\rm
exp})^2}}
\end{equation}
where $B_i^{\rm theor(exp)}$ denote theoretical (experimental) CP-averaged
 branching fractions
and $\Delta B_{i}^{\rm exp}$ is an experimental error for i-th decay channel.
The sum is over all 16 decay channels as in \cite{ZL2,chpZ}.
Experimental branching ratios and their errors are listed in Tables 1 and 2.
The connection between the amplitudes and branching ratios was corrected in our calculations for the lifetime
difference between $B^{+}$ and $B^{0}$:

\begin{equation}
\frac{\tau_{B^{+}}}{\tau_{B^{0}}}=1.068
\end{equation}
\\
We considered two sets of data. The first one was the same as in \cite{ZL2}. The second one was used
in \cite{chpZ}. Data in Table 2 are more recent and differ from the previous ones in a couple of entries.
We performed fits in three general cases:
\begin{enumerate}
\item {without long-distance charming penguin contributions and with $|T|$, $|P|$, $|S'|$, $\gamma$
treated as free parameters}
\item {with long-distance charming penguins described by real $P_{cef}$ as an additional parameter and
$\delta=0$, which is consistent with
 calculations done in \cite{chpZ} but without any assumed connection between $P_{c}$ and $P_{t}$}
\item {with long-distance charming penguins described by two additional free parameters: $\delta$, $P_{cef}$.}
\end{enumerate}

\begin{table}[h]
\caption{Fits to the first set of data (in units of $10^{-6}$)} \label{tab:fit1}
\begin{center}
{\footnotesize
\begin{tabular}{|l|l|c|c|c|c|}
%&&\\
\hline
Decay channel                 & Exp          &SD amplitudes  &\multicolumn{3}{|c|}{Charming penguin}       \\

                             &              &only   &(case 2)    &   \multicolumn{2}{|c|}{(case3)}\\

                            &              & (case 1)  &$\delta=0^{o}$ &$\gamma$ free  &$\gamma=64.5^o$  \\

\hline \hline
$(B^+ \to \pi ^+ \pi ^0)$    &$5.8\pm 1.0$  &$5.01$  &5.65	      &$5.73$    &$5.85$		 \\
$(B^+ \to K ^+ \bar{K}^0)$   &$0.0\pm 2.0$  &$0.68$  &0.71	      &$2.10$    &$1.81$		 \\
$(B^+ \to\pi ^+ \eta)$       &$2.9\pm 1.1$  &$2.15$  &1.76	      &$2.47$    &$2.24$		 \\
$(B^+ \to\pi ^+ \eta ')$     &$0.0\pm 7.0$  &$1.07$  &0.88	      &$1.24$    &$1.12$		 \\
\hline
$(B^0_d \to \pi ^+ \pi ^- )$ &$4.7\pm 0.5$  &$4.90$  &4.78	      &$4.76$    &$4.75$		 \\
$(B^0_d \to\pi ^0 \pi ^0)$   &$1.9\pm 0.7$  &$0.62$  &0.73	      &$1.50$    &$1.36$		 \\
$(B^0_d \to K^+ K^-)$        &$0.0\pm 0.6$  &$0.00$  &0 	      &$0.00$    &$0.00$		 \\ 			    
$(B^0_d \to K^0 \bar{K}^0)$  &$0.0\pm 4.1$  &$0.62$  &0.66	      &$1.94$    &$1.67$		 \\ 	   	    

\hline
$(B^+ \to \pi ^+ K ^0 )$     &$18.1\pm 1.7$ &$18.40$ &19.21	      &$18.67$   &$20.41$		 \\ 	   
$(B^+ \to \pi ^0 K ^+ )$     &$12.7\pm 1.2$ &$13.11$ &13.10	      &$11.61$   &$10.63$		 \\ 	   
$(B^+ \to\eta K^+)$          &$4.1\pm 1.1$  &$2.46 $ &2.30	      &$4.30 $   &$3.96$		 \\ 	   
$(B^+ \to\eta ' K^+)$        &$75\pm 7.0$   &$73.00$ &73.37	      &$68.91$   &$69.69$		 \\ 	   
\hline
$(B^0_d \to\pi ^- K^+)$      &$18.5\pm 1.0$ &$18.76$ &18.60	      &$18.38$   &$18.60$		 \\ 	   
$(B^0_d \to\pi ^0 K^0)$      &$10.2\pm 1.2$ &$6.20$  &6.57	      &$7.76$    &$9.12$		 \\ 	   
$(B^0_d \to\eta K^0)$        &$0.0\pm 9.3$  &$1.81$  &1.79	      &$3.19$    &$4.22$		 \\ 	   
$(B^0_d \to\eta ' K^0)$      &$56\pm 9.0$   &$66.28$ &67.36	      &$62.35$   &$66.12$		 \\ 	   
\hline

$|T|$&                                      &$2.60$  &2.76	      &$2.78$    &$2.81$                 \\

$|P|$&                                      &$0.79$  &1.45	      &$2.59$    &$1.92$                 \\

$|S'|$&                                     &$1.75$  &1.72	      &$2.46$    &$3.02$                 \\

$P_{cef}$&                                  &        &-0.77	      &$-2.81$   &$-2.32$                \\

$\gamma$&                                   &$103^o$ &$94^o$	      &$110^o$   &$64.5^o$               \\

$\delta$&                                   &        &$0^o$	      &$\pm18^o$ &$\pm26^o$              \\

$f_{m}$&                                    &$15.36$ &14.79	      &$6.37$    &$9.39$   \\
\hline
\end{tabular}
}
\end{center}
\end{table}

\begin{table}[h]
\caption{Fits to the second set of data (in units of $10^{-6}$)} \label{tab:fit2}
\begin{center}
{\footnotesize
\begin{tabular}{|l|l|c|c|c|c|}
%&&\\
\hline
Decay channel                & Exp          &SD amplitudes  &\multicolumn{3}{|c|}{Charming penguin}       \\

                             &              &only   &(case 2)    &   \multicolumn{2}{|c|}{(case3)}\\

                            &              & (case 1)  &$\delta=0^{o}$ &$\gamma$ free  &$\gamma=64.5^o$  \\

\hline \hline
$(B^+ \to \pi ^+ \pi ^0)$    &$5.3\pm 0.8$  &$4.27$  &5.32	   &$5.05$    &$5.40$		    \\
$(B^+ \to K ^+ \bar{K}^0)$   &$0.0\pm 2.4$  &$0.69$  &0.96	   &$2.55$    &$1.58$		    \\
$(B^+ \to\pi ^+ \eta)$       &$4.2\pm 0.9$  &$2.66$  &2.04	   &$3.04$    &$2.29$		    \\
$(B^+ \to\pi ^+ \eta ')$     &$0.0\pm 4.5$  &$1.33$  &1.02	   &$1.52$    &$1.14$		    \\
\hline						     		   	     
$(B^0_d \to \pi ^+ \pi ^- )$ &$4.6\pm 0.4$  &$5.09$  &4.76	   &$4.75$    &$4.70$		    \\
$(B^0_d \to\pi ^0 \pi ^0)$   &$1.9\pm 0.5$  &$0.51$  &0.83	   &$1.65$    &$1.18$		    \\
$(B^0_d \to K^+ K^-)$        &$0.0\pm 0.6$  &$0.00$  &0 	   &$0.00$    &$0.00$		    \\
$(B^0_d \to K^0 \bar{K}^0)$  &$0.0\pm 1.8$  &$0.64$  &0.89	   &$2.35$    &$1.46$		    \\
						     		   	     
\hline
$(B^+ \to \pi ^+ K ^0 )$     &$21.8\pm 1.4$ &$19.10$ &22.11	   &$22.44$   &$21.57$  	    \\
$(B^+ \to \pi ^0 K ^+ )$     &$12.8\pm 1.1$ &$11.97$ &12.45	   &$10.92$   &$11.39$  	    \\
$(B^+ \to\eta K^+)$          &$3.2\pm 0.7$  &$2.03 $ &1.57	   &$2.71$    &$3.04$		    \\
$(B^+ \to\eta ' K^+)$        &$77.6\pm 4.6$ &$74.02$ &76.18	   &$75.27$   &$74.64$  	    \\
\hline						     		   	     
$(B^0_d \to\pi ^- K^+)$      &$18.2\pm 0.8$ &$17.57$ &18.20	   &$19.33$   &$19.01$  	    \\
$(B^0_d \to\pi ^0 K^0)$      &$11.9\pm 1.5$ &$6.86$  &8.03	   &$9.86$    &$9.14$		    \\
$(B^0_d \to\eta K^0)$        &$0.0\pm 4.6$  &$1.76$  &1.63	   &$3.85$    &$3.26$		    \\
$(B^0_d \to\eta ' K^0)$      &$65.2\pm 6.0$ &$68.66$ &72.32	   &$73.14$   &$70.76$  	    \\
\hline

$|T|$&                                      &$2.36$  &2.68	   &$2.61$    &$2.7$		    \\
						     		   	     
$|P|$&                                      &$0.83$  &2.06	   &$2.63$    &$1.9$		    \\
						     		   	     
$|S'|$&                                     &$1.77$  &1.69	   &$2.96$    &$2.61$		    \\
						     		   	     
$P_{cef}$&                                  &	     &-1.45	   &$-3.05$   &$-2.07$  	    \\
						     		   	     
$\gamma$&                                   &$85^o$  &$68^o$	   &$22^o$    &$64.5^o$ 	    \\
$\delta$&                                   &	     &$0^o$	   &$\pm19^o$ &$\pm26^o$	    \\

$f_{m}$&                                    &$27.98$ &24.73	   &$14.97$   &$15.71$  	   \\
\hline
\end{tabular}
}
\end{center}
\end{table}

\begin{figure}[h]

\includegraphics[angle=-90,width=0.5\textwidth]{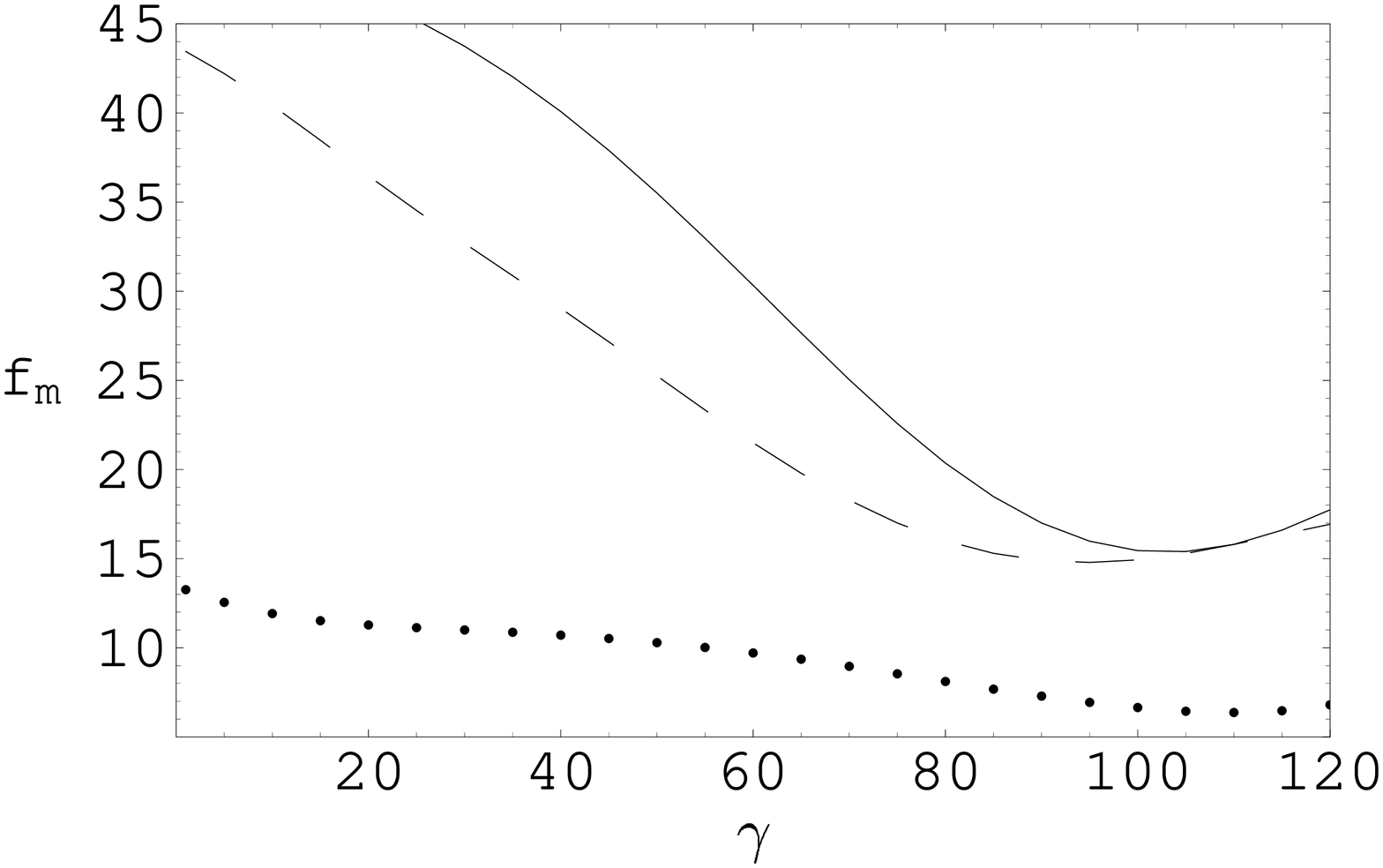}
\includegraphics[angle=-90,width=0.5\textwidth]{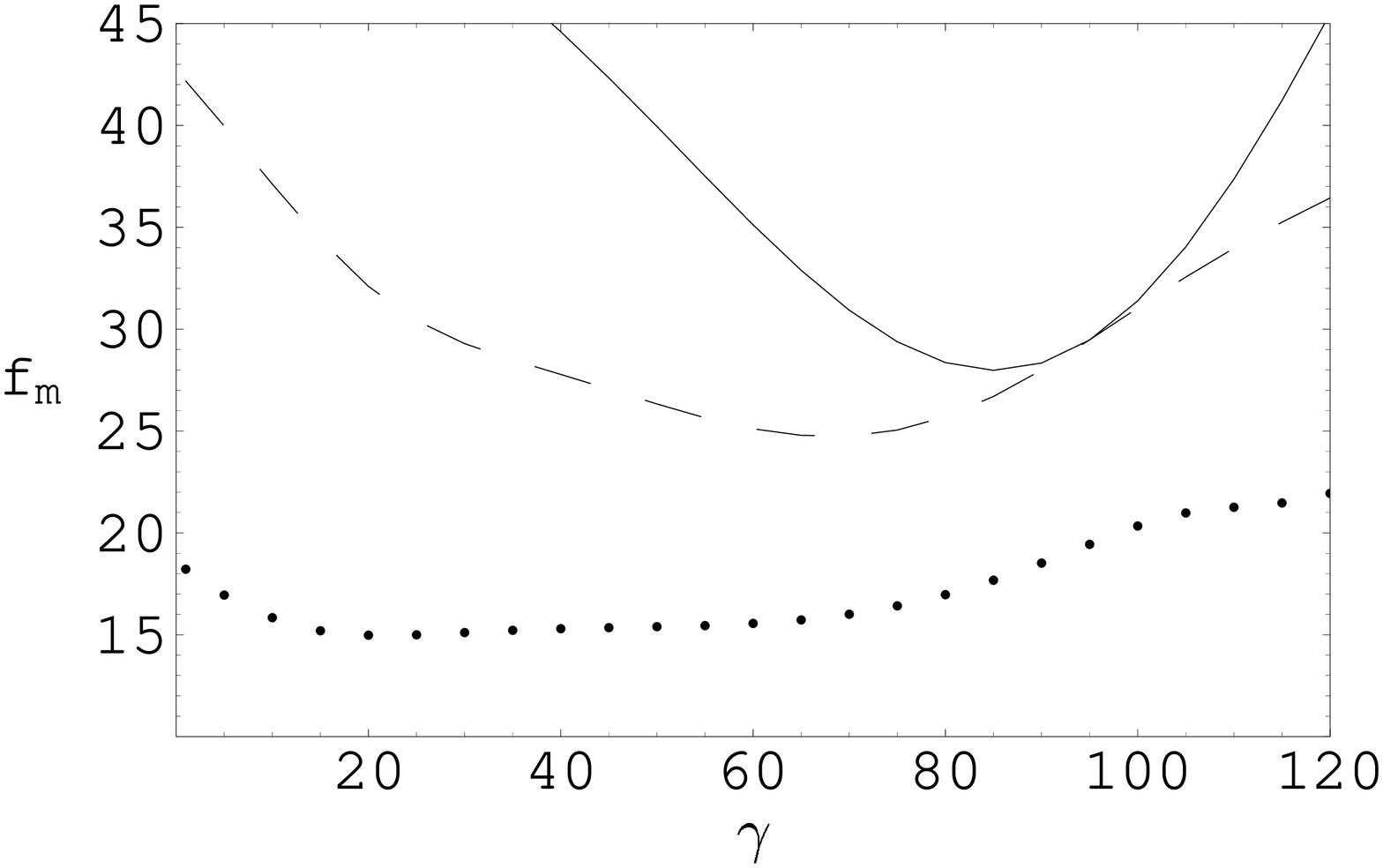}

\caption{\small Dependence of $f_{m}(\gamma)$ on $\gamma$ for the first (left) and second (right) set of data.
Solid lines denote case without charming penguin, dashed lines - case with charming  penguin and $\delta =0$, 
dotted lines - case with charming penguin and $\delta$ let free.}

\end{figure}

Results of the fits are contained in Tables 1, 2. The branching fractions were calculated for the best
fits and for the fit with fixed $\gamma=64.5^{o}$. The minimums $f_{m}$ obtained by minimising $f$ of (10) are showed in the last rows of the tables.
In general, the fitted values of  $\gamma$ are far from the standard model prediction.
To find out what happens one should study the dependence of the fitted function on $\gamma$.
Let us denote by $f_{m}(\gamma)$ the minimum values of $f$ obtained when keeping $\gamma$ fixed.
The function $f_{m}(\gamma)$ is obtained either by setting $P_{cef}$=0, or by assuming $\delta=0$ while letting
 $P_{cef}$ free, or by letting both  $P_{cef}$ and $\delta$ free.
Figure 1 shows $f_{m}(\gamma)$ for the first (left) and second (right) set of data. The worst fits are those
without charming penguins (solid lines). The minimal values were achieved for $\gamma=103^o$ and
$\gamma=85^o$ respectively. For both fits with charming penguin and the strong phase $\delta_{c}=0$ (dashed lines), the
best fit corresponds to $\gamma$
shifted down by  $9^{o}$($17^{o}$)  and a slightly lower value of $f_{m}$. In the third  case shown (dotted
 lines) $\delta$ was let free. For the first set of data, the $f_{m}(\gamma)$ is fairly small over the whole
 region shown $(\gamma \in (0,120^{o}))$.
  For the second, more recent set of data this region is restricted to about $10^{o}-80^{o}$. Since the values of
 $f_{m}$ differ a little in the above-mentioned region we should rather think of an allowed range of $\gamma$.
\\

The values of fitted parameters $|P|$,$|S^{'}|$,$|P_{cef}|$ vary for different values of $\gamma$.
%$|T|$ changes in the range 2,3-2,8(2,5-2,8) for $\gamma$ between $0^{o}$ and $120^{o}$
The most stable are the ratio $\frac{P_{cef}}{P}$ and the strong phase $\delta$. $|\frac{P_{cef}}{P}|$
changes from 1.1 to 1.3(1.2) only. The function $f_{m}(\delta ,\gamma)$ has a deep minimum around
$ \delta \approx \pm 20^{o}$ (Fig.2) for a wide range of fits with fixed $\gamma$. Both positive and negative signs of $\delta$ are allowed
%due to the fact that
as the fitted function is
symmetric under $\delta \leftrightarrow -\delta$. The fact that the ratio $|\frac{P_{cef}}{P}|$ is close to unity  is in agreement with the calculation in \cite{buras}. On the other hand, for the
 best fits with $\delta=0$ the ratio $|\frac{P_{cef}}{P}|$ is about  0.53(0.7). This value for the second set of data is higher than that assumed in \cite{chpZ}.
\begin{figure}[h]
\begin{center}

\includegraphics[width=0.9\textwidth]{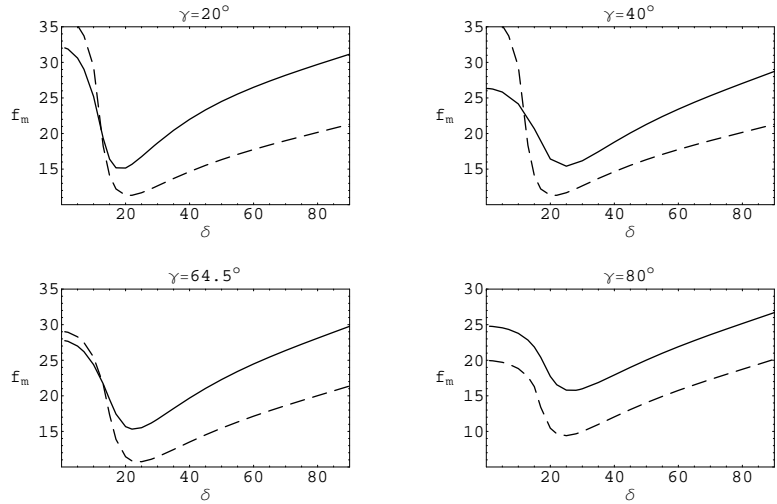}

\caption{\small Dependence of $f_{m}(\gamma ,\delta)$ on $\delta$ for selected
values of $\gamma$, dashed (solid) lines denote the first (second) set of data.}
\end{center}

\end{figure}

\begin{table}[h]
\caption{Asymmetries generated by charming penguin for $\delta>0$ (for $\delta<0$ asymmetries are of opposite
sign)}
\begin{center}
{\footnotesize
\begin{tabular}{|l|c|c|c|c|c|}
%&&\\
\hline
Decay channel &  \multicolumn{2}{|c|}{First set of data}&\multicolumn{2}{|c|}{Second set of data} &Experiment\\
              & $\gamma$ fitted&$\gamma=64.5^o$ &$\gamma$ fitted&$\gamma=64.5^o$                  &        \\
\hline \hline
$(B^+ \to \pi ^+ \pi ^0)$ &$0$ &0              &$0$ &0&$-0.07 \pm 0.14$\\	     								   
$(B^+ \to K ^+ \bar{K}^0)$ &-0.93&-0.90        &-0.90&-0.98&-\\ 		     								   
$(B^+ \to\pi ^+ \eta)$   &0.48&0.87            &0.47&0.76&$-0.44 \pm 0.18 \pm 0.01$\\								   
$(B^+ \to\pi ^+ \eta ')$&0.48&0.87             &0.47&0.76&-\\			     								   
\hline
$(B^+ \to \pi ^+ K ^0 )$&0.11&0.08             &0.04&0.07&$0.02 \pm 0.06$\\	     								   
$(B^+ \to \pi ^0 K ^+ )$ &-0.21&-0.28          &-0.09&-0.23&$0.00 \pm 0.12$\\	     								    
$(B^+ \to\eta K^+)$   &0&0                     &0&0&$-0.52 \pm 0.24 \pm 0.01$\\      								   
$(B^+ \to\eta ' K^+)$ &0.006&-0.004            &0.005&-0.004&$0.02 \pm 0.042$\\      								   
\hline
$(B^0_d \to\pi ^- K^+)$ & -0.19&-0.24          & -0.075&-0.21&$-0.09 \pm0.03$\\      								   

\hline
\end{tabular}
}
\end{center}
\end{table}

Charming penguins with a nonvanishing strong phase may be a source of direct CP asymmetries.
The predicted values were calculated for the same points as in Tables 1,2. The results are
given in Table 3 together with the averages from Belle, BABAR and CLEO experiments \cite{Rosner}.
The main features are large asymmetries in the $\Delta S=0$ sector with relatively small asymmetries
for the $\Delta S=1$ decays channels. We are not able to predict the absolute signs of the asymmetries
since we have two allowed signs of $\delta$. The asymmetry for $(B^+ \to \pi ^+ K ^0 )$
is a pure $\lambda^{4}$ effect and shows a potential influence of this correction.

\section{Conclusions}
Our results permit to draw the following conclusions:
\begin{enumerate}
\item{Even without the charming penguins the value of angle $\gamma$  extracted from the fit depends on the details of data. More recent data prefer the value of $\gamma$ 
more in accordance with the expectations of the standard model.  }
\item{If we admit the non-zero value of the charming penguin (with strong phase equal zero), the fitted values of $\gamma$ may move toward the SM value by $10^{o}-15^{o}$.}
\item{Admitting strong phase of the charming penguin as a free parameter leads to a relatively flat function $f_{m}(\gamma)$ i.e. it allows a wide range of $\gamma$.
 This means that there is probably  too much freedom in the fits. However, the fitted strong phase $\delta$ is
 relatively stable and close to $\pm 20^{o}$.}
\end{enumerate}

{\it Acknowledgements}. I would like to thank P.  \.Zenczykowski for helpful discussions and comments.

%\newpage

\end{document}